\documentclass{PoS}
\usepackage{psfrag,epsfig,graphicx,graphics,xcolor}
\usepackage{wrapfig}
\usepackage{amsmath}

\newcommand{\xb}{{\underline x}}
\newcommand{\eb}{{\underline e}}

\newcommand{\kb}{{\underline k}}

\newcommand{\be}{\begin{equation}}
\newcommand{\ee}{\end{equation}}
\newcommand{\eq}{\end{equation}}

\newcommand{\qb}{\bar{q}}

\title{Theory and phenomenology of helicity amplitudes for high energy exclusive leptoproduction of the $\rho-$meson}

\ShortTitle{Th. and pheno. of helicity amplitudes for h. e. exclusive leptoproduction of the $\rho-$meson}

\author{I.~V.~Anikin\\
       Bogoliubov Laboratory of Theoretical Physics, JINR, 141980 Dubna, Russia\\
       Institute for Theoretical Physics, University of Regensburg, D-93040 Regensburg, Germany\\
       E-mail: \email{anikin@theor.jinr.ru}}

\author{\speaker{A.~Besse}\\
        LPT, Universit\'e Paris-Sud, CNRS, 91405, Orsay, France\\
        National Center for Nuclear Research (NCBJ), Warsaw, Poland\\
        E-mail: \email{adrien.besse@th.u-psud.fr}}

\author{D.Yu.~Ivanov\\
       Sobolev Institute of Mathematics and Novosibirsk State University, 630090 Novosibirsk, Russia\\
       E-mail: \email{d-ivanov@math.nsc.ru}}

\author{B.~Pire\\
       CPHT, {\'E}cole Polytechnique, CNRS, 91128 Palaiseau Cedex, France\\
       E-mail: \email{Bernard.Pire@cpht.polytechnique.fr}}

\author{L. Szymanowski\\
       National Center for Nuclear Research (NCBJ), Warsaw, Poland\\
       E-mail: \email{Lech.Szymanowski@fuw.edu.pl}}

\author{S. Wallon\\
       LPT, Universit\'e Paris-Sud, CNRS, 91405, Orsay, France\\
       UPMC Univ. Paris 06, facult\'e de physique, 4 place Jussieu, 75252 Paris Cedex 05,
France\\
       E-mail: \email{Samuel.Wallon@th.u-psud.fr}}

\abstract{We review here two approaches to describe hard leptoproduction of transversally polarized $\rho-$meson, based on recent calculation of the $\gamma^*_T \to \rho_T$ impact factor up to twist~3 accuracy in the collinear factorization frame, including 2- and 3- particles Fock-states. 
The first approach uses a model of the unintegrated gluon density (the proton impact factor) which allows a comparison of our predictions with H1 and ZEUS data for the ratios of helicity amplitudes $T(\gamma^{*}_T \to \rho_T)/T(\gamma^{*}_L \to \rho_L)$ and $T(\gamma^{*}_T \to \rho_L)/T(\gamma^{*}_L \to \rho_L)$. In the second approach, we transform the $\gamma^*_T \to \rho_T$ impact factor into the impact parameter space. 
We show that the transformed amplitude factorizes according to conventional dipole picture. We shortly discuss a way to implement the nucleon saturation effects in our approach.}

\FullConference{Sixth International Conference on Quarks and Nuclear Physics,\\
		April 16-20, 2012\\
		Ecole Polytechnique, Palaiseau, France}

\begin{document}
\section{Helicity amplitudes of the hard leptoproduction of the $\rho-$meson}
\label{chapitre}
\begin{figure}[htb]
\centering
\psfrag{rho}[cc][cc]{$\hspace{.8cm}\rho (p_{\rho})$}
\psfrag{k}[cc][cc]{$k$}
\psfrag{rmk}[cc][cc]{}
\psfrag{l}[cc][cc]{}
\psfrag{q}[cc][cc]{}
\psfrag{lm}[cc][cc]{\raisebox{.6cm
}{$\quad \quad \, \, \, \Gamma \ \,\quad \Gamma$}}
\psfrag{H}[cc][cc]{
$ H_{q \bar{q}}$}
\psfrag{S}[cc][cc]{
$ S_{q \bar{q}}$}
\begin{tabular}{ccc}
\hspace{-1.5cm}\epsfig{file=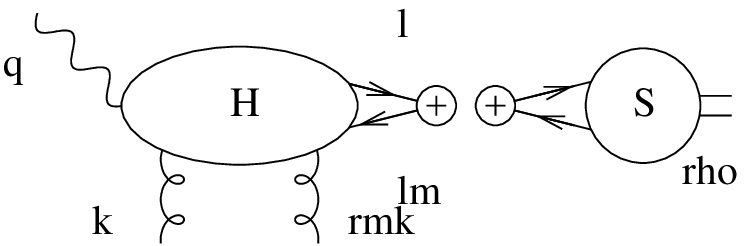,width=.45\linewidth}&\raisebox{4
    \totalheight}{\hspace{1cm}+}&
\psfrag{H}[cc][cc]{
 $\partial_{\perp} H_{q \bar{q}}$}
\psfrag{S}[cc][cc]{
$ \partial_\perp S_{ q \bar{q}}$}
\epsfig{file=FiertzHSqq_rhofact.eps,width=.43\linewidth}\\
&\raisebox{3.5
    \totalheight}{\hspace{-.4\linewidth}+}
  &
    \psfrag{k}[cc][cc]{}
\psfrag{rmk}[cc][cc]{}
\psfrag{l}[cc][cc]{}
\psfrag{q}[cc][cc]{}
\psfrag{Hg}[cc][cc]{
$H_{q \bar{q}g}$}
\psfrag{Sg}[cc][cc]{
$\!S_{q \bar{q}g}$}
\psfrag{H}[cc][cc]{
$H_{q \bar{q}g}$}
\psfrag{S}[cc][cc]{
$ S_{q \bar{q}}$}
\psfrag{S}[cc][cc]{
$ S_{ q \bar{q}g}$}
\psfrag{lm}[cc][cc]
{\raisebox{.2cm
}{$\quad \,\,\,\, \, \, \, \Gamma \ \, \, \Gamma$}}
\hspace{-.4\linewidth}\epsfig{file=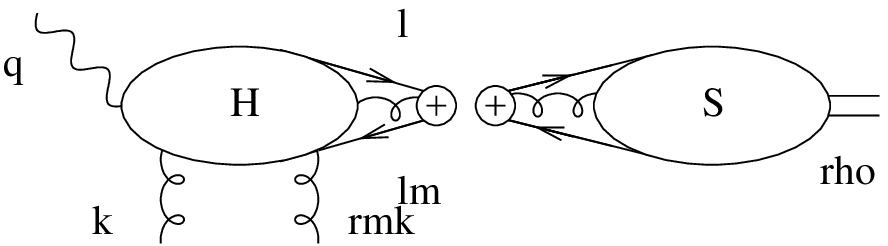,width=.45\linewidth}
\end{tabular}
\caption{Illustration of the 3 different types of terms entering the computation up to twist~3. From left to right: the 2-parton collinear contribution, the 2-parton first order contribution in $\ell_{\perp} $, the collinear term of the 3-parton contribution. $\Gamma$ stands for the set of Dirac matrices inserted using Fierz identity in the spinor space.}
\label{fig1}
\end{figure}

\begin{figure}[ht!]
\begin{center}
\includegraphics[width=7cm]{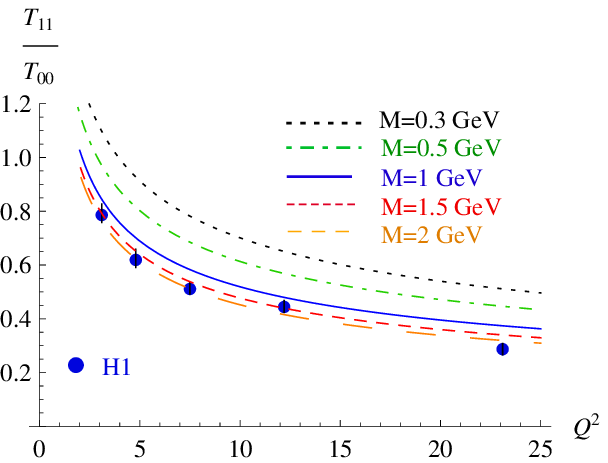} \includegraphics[width=7cm]{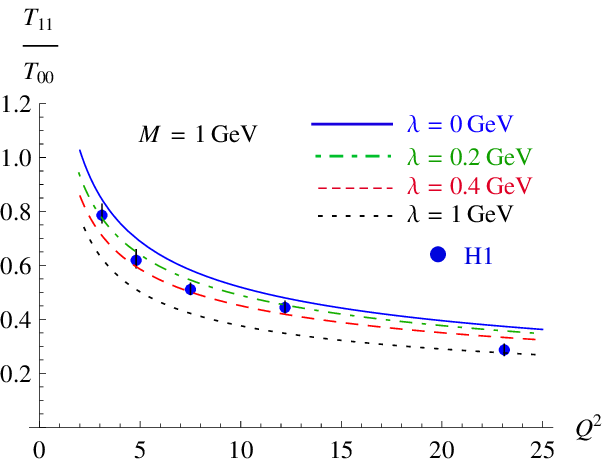}
\caption{Predictions for the ratio $T_{11}/T_{00}$ as a function of $Q^2$, compared to the data from H1~\cite{H1}.
Left: fixed $\lambda = 0$, and various values for $M$. Right: fixed scale M = $1\:$GeV, and various values of $\lambda$.}
\label{fig1.2}
\end{center}
\end{figure}

In the Regge inspired factorization scheme, helicity amplitudes of the hard diffractive $\rho-$meson production 
\begin{displaymath}
\gamma^{*}(q,\lambda_{\gamma}) \, N(p_2) \rightarrow \rho (p_{\rho},\lambda_{\rho}) \, N(p_2)
\end{displaymath}
 are expressed in terms of the $\gamma^*(\lambda_{\gamma})\to \rho(\lambda_{\rho})$ impact factor ($\Phi^{\gamma^{*}(\lambda_{\gamma}) \rightarrow \rho(\lambda_{\rho})}$) and the corresponding nucleon impact factor ($\Phi^{N \rightarrow N}$).
At two gluon (Born) approximation, the helicity amplitudes read 
\begin{eqnarray}
\label{defImpactRep}
 T_{\lambda_{\rho}\lambda_{\gamma}}(\underline{r};Q , M)= is\int \frac{d^2\kb}{(2\pi)^2}\frac{1}{\kb^2(\kb-\underline{r})^2}\,\Phi^{N \rightarrow N} (\kb,\underline{r};M^2)\,\Phi^{\gamma^{*}(\lambda_{\gamma}) \rightarrow \rho(\lambda_{\rho})}(\kb,\underline{r};Q^2)\,,
\label{eqT}
 \end{eqnarray}
(using underlined letters for the euclideans two dimensional transverse vectors).
The momenta $q$ and $p_{\rho}$ are parameterized using the Sudakov decomposition in terms of two light cone vectors $p_1$ and $p_2$ as $q=p_1-\frac{Q^2}{s}p_2$ and $p_{\rho}\equiv p_1+\frac{m_\rho^2-t+t_{min}}{s}p_2+r_{\perp}$, with $2 p_1 \cdot p_2=s $ and $q^2=-Q^2$ the virtuality of the photon.
The computation of the $\gamma^{*}(\lambda_{\gamma})  \rightarrow \rho (\lambda_{\rho}) $ impact factor is performed within the collinear factorization of QCD. 
The dominant contribution $\gamma^{*}_L  \rightarrow \rho_L$ transition (twist~2) has been computed long time ago~\cite{GinzburgPanfilSerbo} while a consistent treatment of the twist~3 $ \gamma^{*}_T  \rightarrow \rho_T$ transition has been performed only recently in references~\cite{Anikin2009,Anikin2010}. 
Impact factors involve a hard part where the hard photon decays into partons that interact with the $t-$channel gluons and soft parts where these partons hadronize into a $\rho-$meson. Soft and hard parts are factorized in momentum space by expanding the hard parts around the longitudinal components of the momenta of the partons
, collinear to the direction of the $\rho-$meson momentum. Fierz identity applied to spinor and color spaces, is used to factorize color and spinor indices linking hard and soft parts. 
Up to twist~3, the amplitude involves the 2- and 3-particle correlators, which enter in the soft part, and the corresponding hard part, as illustrated in Fig.~\ref{fig1}.
Soft parts correspond to the Fourier transforms of the vacuum to $\rho-$meson matrix elements that are parameterized by a set of five 2-parton distribution amplitudes (DAs) $\{  \varphi_1,  \varphi_A, \varphi_3,  \varphi_{1T}, \varphi_{AT} \}$ and two 3-parton DAs $\{  B(y_1,y_2), D(y_1,y_2)\}$. Due to the equation of motion of QCD and the $n-$independence condition (cf.~\cite{Anikin2010} for details), the DAs $\{  \varphi_A, \varphi_3,  \varphi_{1T}, \varphi_{AT} \}$ are expressed in terms of the 2-parton DA $\varphi_1$ (Wandzura-Wilczek (WW) contributions) and of the two 3-parton DAs $\{  B(y_1,y_2), D(y_1,y_2)\}$ (the genuine 3-parton contribution). 
In our model we take into account effects of the ERBL evolution involving the factorization scale $\mu_F$ dependence. 
 In our phenomenological estimate~\cite{Anikin2011} we have chosen a simple model proposed by Gunion and Soper in Ref.~\cite{GunionSoper} for the proton impact factor, which depends on two parameters $M$ and $A$,
\begin{displaymath}
\label{ProtonIF}
\Phi_{N \to N}(\kb,\underline{r};M^2)\!\!=\!\! A \delta_{ab}\!\!\left[\frac{1}{M^2+(\frac{\underline{r}}{2})^2}-\frac{1}{M^2+(\kb-\frac{\underline{r}}{2})^2}\right]\!\!.
\end{displaymath}
Fig.~\ref{fig1.2} shows the results for the ratios of the helicity amplitudes that we obtain within our approach, as a function of the parameter $M$ and of an infra-red cut off $\lambda$ for the integral over the transverse momenta of the $t-$channel gluons. Our predictions are compared to the data of H1~\cite{H1}. 
They are in fairly good agreement with the data for a value of $M$ between $0.9\:$GeV and $5\:$GeV and they weakly depend on the value of $M$ and $\mu_F$. The dependence on the cut-off $\lambda$ shows that soft gluons with momenta smaller than $\Lambda_{QCD}$ ($\left|\kb\right|\leq \Lambda_{QCD}$) give a small contribution to the final result while the contribution of the gluons with $\left|\kb\right|\sim 1\:$GeV cannot be neglected and thus calls for an inclusion of the saturation effects in the nucleon.
\section{Dipole representation of the scattering amplitude $\gamma^*_T \,N\,\to\, \rho_T\,N$ up to twist~3}
One can expect, that at high energies the diffractive leptoproduction of the $\rho-$meson can be represented within the dipole picture introduced in Ref.~\cite{Mueller,NikZak}, where the high energy scattering amplitude factorizes on the one hand side into the initial and final wave functions of the projectile $\gamma^*$ and the outgoing $\rho-$meson, and on the other side into the color dipole scattering amplitude on the nucleon. 
In Ref.~\cite{Besse2012}, we show that within the collinear approximation up to twist~3 it is indeed the case. 


The impact factor in the impact parameter representation, i.e. expressed in terms of the Fourier transforms with respect to the transverse momenta of the partons of the hard and soft parts, can be written in the form (Ref.~\cite{Besse2012}),
\begin{align}
 \Phi^{\gamma^*_T \to \rho_T}&=-\frac{i}{4}m_{\rho} f_{\rho} \int dy \int \frac{d^2 \xb}{(2 \pi)}\left\{ \tilde{H}_{q\qb}^{\gamma, \mu}(y,\xb)\left(-i\varphi_3(y) e_{\rho \mu}^*+\varphi_{1T}(y) p_{1\mu}(\eb_{\rho}^* \cdot \xb)\right) \right.\nonumber\\
&\left.+\tilde{H}_{q\qb}^{\gamma_5\gamma, \mu}(y,\xb)\left(\varphi_A(y)\varepsilon_{\mu e^*_{\rho} p_1 n}-i\varphi_{AT}(y) p_{1\mu}\varepsilon_{x_{\perp} e^*_{\rho}p_1 n}\right)\right. \nonumber\\
&\left.+\int_0^{y} d y' \int \frac{d^2 \xb'}{(2\pi)^2}\left( \zeta^{V}_{3\rho} B(y',y)   p_{\mu} e_{\rho \perp \alpha} \,\tilde{H}_{q\qb g}^{\alpha,\gamma^{\mu}}(y',y,\xb',\xb)\right.\right.\nonumber\\
&\left.\left.\hspace{2cm}+\zeta^{A}_{3\rho} i D(y',y) p_{\mu} \varepsilon_{\alpha e_{\rho \perp} p n}\,  \tilde{H}_{q\qb g}^{\alpha,\gamma^{\mu}\gamma_5}(y',y,\xb',\xb)\right) \right\}\,,
\label{eqww}
\end{align}
where the set $\{\tilde{H}_{q\qb}^{\gamma, \mu},\tilde{H}_{q\qb}^{\gamma_5\gamma, \mu}\}$ and $\{\tilde{H}_{q\qb g}^{\alpha,\gamma^{\mu}} ,\tilde{H}_{q\qb g}^{\alpha,\gamma^{\mu}\gamma_5}\}$ are corresponding  transforms of the 2-parton and 3-parton hard parts. 

We show further that  the impact factor Eq.~(\ref{eqww}) can be rewritten in the factorized form as
\begin{align}
\label{A-final-non-flip}
\Phi^{\gamma^*_T \to \rho_T}&\propto \frac{m_{\rho}\, f_{\rho}}{\sqrt{2}}\int dy\ \int d^2\xb\, g^2 \,\delta^{ab}\, N(\xb,\kb)\, \psi^{\gamma^*_T\to\rho_T}(y,\xb) \,,
\end{align}
in which we explicitly factorized $g^2 \,\delta^{ab}\, N(\xb,\kb)$ which is the scattering amplitude of a dipole of transverse size $\left| \xb \right|$ interacting with the gluons in the $t-$channel. 
The factorized form in  Eq.~(\ref{A-final-non-flip}) is a rather non trivial result as it appears only after using the QCD equations of motion. 
The expression of  $\psi^{\gamma^*_T\to\rho_T}(y,\xb)$ in Eq.~(\ref{A-final-non-flip}) is the overlap of the wave function of the tranversally polarized photon $ \Psi_{\lambda{\gamma} (\lambda)}^{\gamma^*_{T}}$~\cite{Mueller,NikZak,IvanovWusthoff,GiesekeQiao} and the relevant combinations of DAs $\phi_{\lambda_{\rho} (\lambda)}^{WW}$ of the transversally polarized $\rho-$meson. The function  $\psi^{\gamma^*_T\to\rho_T}(y,\xb)$ reads
\begin{align}
\label{eql}
\psi^{\gamma^*_T\to\rho_T}(y,\xb)&=\sum_{(\lambda)}  \phi_{\lambda_{\rho} (\lambda)}^{WW}(y)
\, \Psi_{\lambda{\gamma} (\lambda)}^{\gamma^*_{T}}(y,\xb) \,,
\end{align}
where $\lambda=\pm \frac{1}{2}$ denotes the helicity of the exchanged quark and 
\begin{align}
 \phi_{\lambda_{\rho} (\lambda)}^{WW}(y)&=-i (\eb^{\lambda_{\rho} *}\cdot\xb)\frac{1}{8 N_c} (\varphi_{AT}(y)+\lambda_{\rho} \,(2\lambda)\, \varphi_{1T}(y))\,.
\end{align}
Beyond the WW approximation, it is necessary to evaluate also the double Fourier transforms of the 3-parton hard parts (in the two last lines in Eq.~(\ref{eqww})). We have shown that they can be reduced to a single Fourier transform consistent with Eq.~(\ref{A-final-non-flip}) by appropriate identification of effective dipole formed by quarks and gluon entering the $\rho-$meson vertex. 
A clear interpretation of this part of $\psi^{\gamma^*_T\to\rho_T}(y,\xb)$ (beyond the WW approximation) is still required, as the 3-parton wave function of the transversally polarized photon is still unknown.

%

The factorized form of Eq.~(\ref{A-final-non-flip}) can be considered as a starting point for including saturation effects in the leptoproduction of the $\rho-$meson, since according to the discussion at the end of Chap.~\ref{chapitre} and Ref.~\cite{Anikin2011} they can lead to important effects.
\section{Conclusion}
We have first shown~\cite{Anikin2011} that a model \`a la Gunion and Soper gives a good description of the  H1 data, although saturation effects has to be incorporated. We then proved in Ref.~\cite{Besse2012} that the twist~3 impact factor of the transition $\gamma^*_T\to\rho_T$ is consistent with the conventionnal color dipole picture.  
We leave for a future work the extension of our approach to the nonforward kinematics which will permit, in particular, to investigate saturation effects at fixed impact parameter, which may be crucial for future collider experiments Ref.~\cite{EIC,LHeC}. 
%

\section{Acknowledgements}

We thank K. Golec-Biernat and L. Motyka for stimulating discussions. This work is supported in part by the Polish NCN grant DEC-2011/01/B/ST2/03915, by the French-Polish Collaboration Agreement Polonium, by the Russian grant RFBR-11-02-00242, by the P2IO consortium and
the Joint Research Activity  ``Study of Strongly Interacting
Matter'' (acronym HadronPhysics3, Grant Agreement
n.283286) under the Seventh Framework Programme
of the European Community.

\end{document}